# The Distributed Computing Paradigms: P2P, Grid, Cluster, Cloud, and Jungle

Brijender Kahanwal[*]
*Assistant Professor, CSE Department, Galaxy Global Group of Institutions, Dinarpur, Ambala, Haryana*
(imkahanwal@gmail.com)
INDIA

Tejinder Pal Singh
*Assistant Professor, Applied Sciences Department, RPIIT, Bastara, Karnal, Haryana*
(tps5675@gmail.com)
INDIA

*Abstract:* The distributed computing is done on many systems to solve a large scale problem. The growing of high-speed broadband networks in developed and developing countries, the continual increase in computing power, and the rapid growth of the Internet have changed the way. In it the society manages information and information services. Historically, the state of computing has gone through a series of platform and environmental changes. Distributed computing holds great assurance for using computer systems effectively. As a result, supercomputer sites and datacenters have changed from providing high performance floating point computing capabilities to concurrently servicing huge number of requests from billions of users. The distributed computing system uses multiple computers to solve large-scale problems over the Internet. It becomes data-intensive and network-centric. The applications of distributed computing have become increasingly wide-spread.

In distributed computing, the main stress is on the large scale resource sharing and always goes for the best performance. In this article, we have reviewed the work done in the area of distributed computing paradigms. The main stress is on the evolving area of cloud computing.

## 1  Introduction

The growing popularity of the Internet and the availability of powerful computers and high-speed networks as low-cost commodity components are changing the way we do computing. Distributed computing has been an essential component of scientific computing for decades. It consists of a set of processes that cooperate to achieve a common specific goal. It is widely recognized that Information and Communication Technologies (ICTs) have revolutionized the everyday practice. Social networks represent a stepping stone in the on-going process of using the Internet to enable the social manipulation of information and culture. Mostly social network sites are implemented on the concept of large distributed computing systems. These are running in centrally controlled data centers. However, the trend in these massively scalable systems is toward the use of peer-to-peer, utility, cluster, and jungle computing. The utility computing is basically the grid computing and the cloud computing which is the recent topic of research. This classification is well shown in the Figure 1.1.

With the increasing heterogeneity of the underlying hardware, the efficient mapping of computational problems onto the 'bare metal' has become vastly more complex. There are many challenges of distributed computing as follows:

*Transparency* means to hide distribution from the users at the high levels and to hide the distribution from the programs at the low levels. There are more forms of transparency as Location, Migration, Replication, Concurrency, and Parallelism. *Flexibility* should be easy to develop. *Reliability* encompasses some factors like no data loss, secure system, and fault tolerant systems. *Performance* should be high. *Scalability* should scale indefinitely.

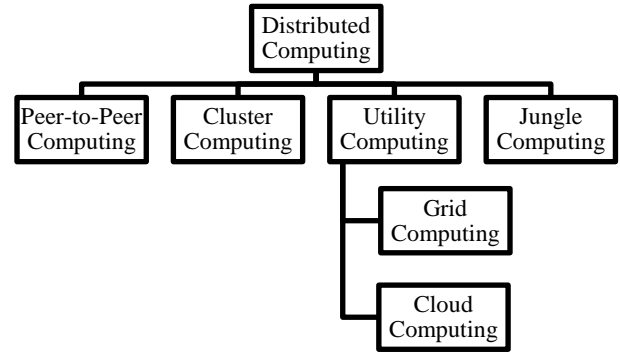

Figure 1.1:   Classification of Distributed Computing

This review paper covers the distributing technologies. In the section 3[rd] peer-to-peer computing is elaborated; in section 4[th], the cluster computing; in section 5[th] utility computing which has the subsections about grid computing and the cloud computing; and in section 6[th], the jungle computing. This paper gives a good introductory knowledge about the distributing computing.

## 2  Related Works

The computing industry is one of the fastest growing industries and it is stimulated by the rapid technological developments in the areas of computer hardware and software. The technological advances in hardware include chip development and fabrication technologies, fast and cheap microprocessors, as well as high bandwidth and low latency interconnection networks. Among them, the recent advances in electronics technology have played a major role in the development of powerful sequential and parallel computers.

Software technology is also developing fast. Mature software, such as Operating Systems, programming languages, development methodologies, and tools, are now available. This has enabled the development and deployment of applications catering to scientific, engineering, and commercial needs. It should also be noted that grand challenging applications, such as weather forecasting and earthquake analysis, have become the main driving force behind the development of powerful parallel computers.

Distributed systems can be considered conventional networks of independent computers. They have multiple system images, as each node runs its own operating system, and the individual machines in a distributed system could be, for example, combinations of Massively Parallel Processors (MPPs), Symmetric Multiprocessors (SMPs), clusters, and individual computers.

Cloud services are mainly divided into three services delivery models: SaaS (Software as a Service, e.g. Google Mail), PaaS (Platform as a Service, e.g. Google AppEngine) and IaaS (Infrastructure as a Service, e.g. Amazon EC2). Since the work presented in this chapter is strongly related to Infrastructure as a Service model, we only focus in this section on this category of service. IaaS providers aim to offer resources to users as pay-as-you-go manner. A key provider of such a service is Amazon through its Elastic Cloud Computing (EC2) and Simple Storage Service (S3).

The academicians and the giant groups are doing their best to comeback with the new concepts of the distributing computing and they have given so many good results. But there is always an intension to develop a better technology, so we are here and cloud computing is the recent topic on which work is in progress.

## 3  Peer-to-Peer Computing

Peer-to-peer (P2P) networking has been working primarily on the scalability issues inherent in distributing resources over a large number of networked processes. In a P2P system, every node acts as both a client and a server, providing part of the system resources. Peer machines are simply client computers connected to the Internet. All client machines act autonomously to join or leave the system freely. This implies that no master-slave relationship exists among the peers. No central coordination or no central database is needed. In other words, no peer machine has a global view of the entire P2P system. The system is self-organizing with distributed control as shown in the Figure 3.1.

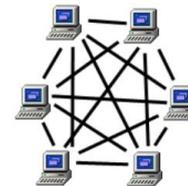

Figure 3.1:    P2P Network

## 4  Cluster Computing

A cluster computing comprises a set of independent or stand-alone computers and a network interconnecting them. It works cooperatively together as a single integrated computing resource. A cluster is local in that all of its component subsystems are supervised within a single administrative domain, usually residing in a single room and managed as a single computer system. The components of a cluster are connected to each other through fast local area networks. To handle heavy workload with large datasets, clustered computer systems have demonstrated impressive results in the past. The architecture of the cluster computing environment is shown in the Figure 4.1.

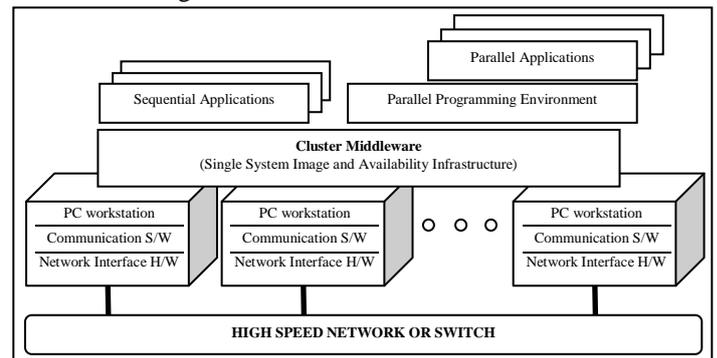

Figure 4.1:    Cluster Computer Architecture

Components of Cluster Computing:
There are so many components of the cluster computing as follows:
o High Performance Computers like PCs, Workstations etc.
o Micro- kernel based operating systems.
o High speed networks or switches like Gigabit Ethernets.
o NICs (Network Interface Cards)
o Fast Communication Protocols and Services
o Cluster Middleware which is hardware, Operating system kernels, applications and subsystems.
o Parallel Programming Environment Tools like compilers, parallel virtual machines etc.
o Sequential and Parallel applications

The cluster middleware is very much capable for offering an elusive and a unified system image.

There is the classification of clusters as shown in the Figure 4.2.

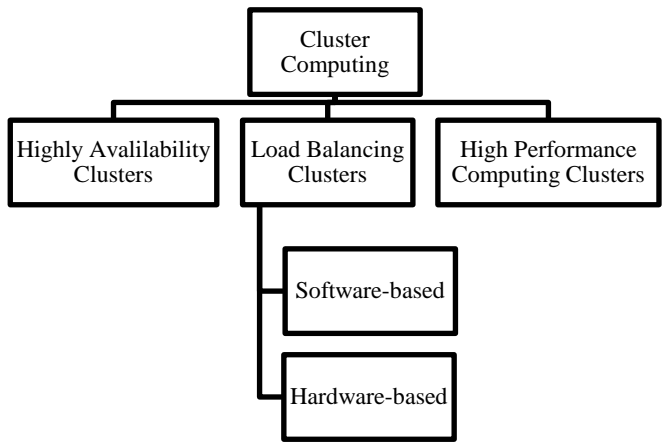

Figure 4.2: Cluster Computing Classification

**High-availability clusters**

These are also called Failover Clusters. These are groups of computers that support server applications. These can be reliably utilized with a minimum of down-time. They operate by harnessing redundant computers in groups or clusters that provide continued service when system components fail. Without clustering, if a server running a particular application crashes, the application will be unavailable until the crashed server is fixed. Such type of clusters remedies this situation by detecting hardware/software faults, and immediately restarting the application on another system without requiring administrative intervention, a process known as failover. As part of this process, clustering software may configure the node before starting the application on it.

**Load balancing clusters**

Load balancing is often required when building solutions that handle large volumes of client requests or that have high demands on security and redundancy. Clusters support multiuser and multitasking environments. These factors accompanied by the heterogeneous nature of the cluster hardware and software leads to the situation that the overall distribution of the workload of a cluster is hard to predict at any particular moment. The static approach to I/O planning is consequently almost useless. The two major categories of load balancing implementations are:

*Software-based load balancing* consists of special software that is installed on the servers in a load-balanced cluster. The software dispatches or accepts requests from the client to the servers, based on different algorithms. The algorithms can be a simple round-robin algorithm or a much more complicated algorithm that considers server affinity. For example, Microsoft Network Load Balancing is a load balancing software for Web farms, and Microsoft Component Load Balancing is a load balancing software for application farms.

*Hardware-based load balancing* consists of a specialized switch or router with software to give it load balancing functionality. This solution integrates switching and load balancing into a single device, which reduces the amount of extra hardware that is required to implement load balancing. Combining the two functions, however, also makes the device more difficult to troubleshoot.

**High Performance Computing Clusters**

The enterprises are now using large-scale clusters that are often shared across departments with easy public access. High performance clusters are used where time to solution is important. They are also used in cases where a problem is so big it can't "fit" on one single computer. To increase computing throughput, HPC clusters are used in a variety of ways.

The easiest way is to allow the cluster to act as a compute farm. Instead of running a job on a local workstation, it is submitted to the cluster for execution. The cluster will manage the resources needed for the job and assign the job to a work queue. When the resource (for instance, a server) is available, the job gets executed and the results are returned to the user. Users who need to run many similar jobs with different parameters or data sets find clusters ideal for this kind of work. They can submit hundreds of jobs and allow the cluster to manage the work flow. Depending on the resources, all the jobs may run at the same time or some may wait in the queue while other jobs finish. This type of computing is local to a cluster node, which means the node doesn't communicate with other nodes, but may need high speed file system access.

## 5 Utility Computing

Utility computing is envisioned to be the next generation of Information Technology evolution that depicts how computing needs of users can be fulfilled in the future IT industry. Its analogy is derived from the real world where service providers maintain and supply utility services, such as electrical power, gas, and water to consumers. Consumers in turn pay service providers based on their usage. Therefore, the underlying design of utility computing is based on a service provisioning model, where users (consumers) pay providers for using computing power only when they need to. Utility computing focuses on a business model, by which customers receive computing resources from a paid service provider. All grid/cloud platforms are regarded as utility service providers. However, cloud computing offers a broader concept than utility computing.

### 5.1 Grid Computing

The aim of Grid computing is to enable coordinated resource sharing and problem solving in dynamic, multi-institutional virtual organizations.

As an electric-utility power grid, a computing grid offers an infrastructure that couples computers, software/middleware, special instruments, and people and sensors together. Grid is often constructed across LAN, WAN, or Internet backbone networks at regional, national, or global scales. Enterprises or organizations present grids as integrated computing resources. They can be viewed also as virtual platforms to support virtual organizations. The computers used in a grid are primarily workstations, servers, clusters, and supercomputers. Personal computers, laptops and PDAs can be used as access devices to a grid system. The grids can be of many types as; Knowledge, Data, Computational, Application Service Provisioning, Interaction or Utility.

These have many pros and cons. Pros are like; these are capable to solve larger, more complex problems in a shorter time, these are easier to collaborate with other organizations, and these make better use of existing hardware. Cons are like; Grid software and standards are still evolving, learning curve to get started, and non-interactive job submission.

## 5.2 Cloud Computing

Cloud computing is another form of utility computing. It is a new term in the computing world and it signals the advent of a new computing paradigm. This new paradigm is quickly developing and attracts a number of customers and vendors alike. The quick development of cloud computing is being fuelled by the emerging computing technologies which allows for reasonably priced use of computing infrastructures and mass storage capabilities. It also removes the need for heavy upfront investment in Information Technology (IT) infrastructure.

Cloud computing is a computing paradigm that involves outsourcing of computing resources with the capabilities of expendable resource scalability, on-demand provisioning with little or no up-front IT infrastructure investment costs. Cloud computing offers its benefits through three types of service or delivery models namely infrastructure-as-a-service (IaaS), platform-as-a-service (PaaS) and software-as-a-Service (SaaS) as shown in the Figure 5.1.

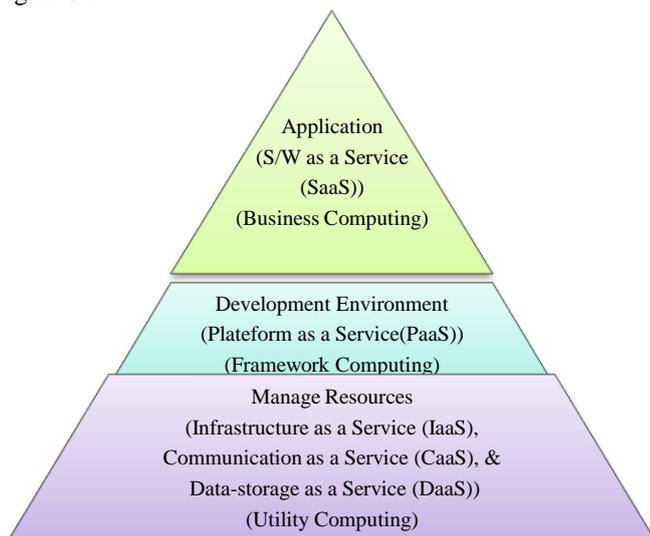

Figure 5.1: Basic concept of cloud computing model and services

It also delivers its service through four deployment models namely, public cloud, private cloud, community cloud and hybrid cloud as the classification is shown in the Figure 5.2.

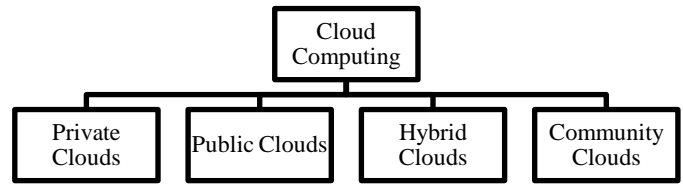

Figure 5.2: Classification of Cloud Computing

**Public clouds** in this deployment the cloud infrastructure is accessible to general public and shared in a pay as you go model of payment. The cloud resources are accessible via the internet and the provider is responsible for ensuring the economies of scale and the management of the shared infrastructure. In this model clients can choose security level they need, and negotiate for service levels. Amazon Web Services EC2 is a public cloud. It is accessible to the general public.

**Private clouds** are another deployment model for cloud services. In this model the cloud resources are not shared by unknown third parties. The cloud resources in this model may be located within the client organization premises or offsite. In this model the client security and compliance requirements are not affected though this offering does not bring the benefits associated with reduced capital expenditure in IT infrastructure investments. In this type of cloud the general public does not have access to the private cloud neither does the organization use the public cloud.

**Hybrid clouds** as its name implies is a model of deployment which combines different clouds for example the private and public clouds. In this model the combined clouds retains their identities but are bound together by standardized technology. In this type of cloud the general public does not have access to the cloud, but the organization uses infrastructure in both the public and private cloud.

**Community clouds** are the fourth deployment model that can be used to deliver cloud computing services. In this model the cloud infrastructure is shared by multiple organizations or institutions that have a shared concern or interest such as compliance considerations, security requirements. This type of cloud may be managed by the organization or by a third party and may be located on-premises or off-premises. In this type of cloud both the public and the organizations forming the community cloud have access to the cloud services offered by the community cloud.

## 6 Jungle Computing

Jungle computing is a simultaneous combination of heterogeneous, hierarchical, and distributed computing resources. In many realistic scientific research areas, domain experts are being forced into concurrent use of multiple clusters, grids, clouds, desktop grids, independent computers, and more. Jungle computing refers to the use of diverse, distributed and highly non-uniform high performance computer systems to achieve peak performance. These new distributed computing paradigms have led to a diverse collection of resources available to research scientists, including stand-alone machines, cluster systems, grids, clouds, desktop grids, etc. as shown in the Figure 6.1 and this varied collection is named as jungle computing.

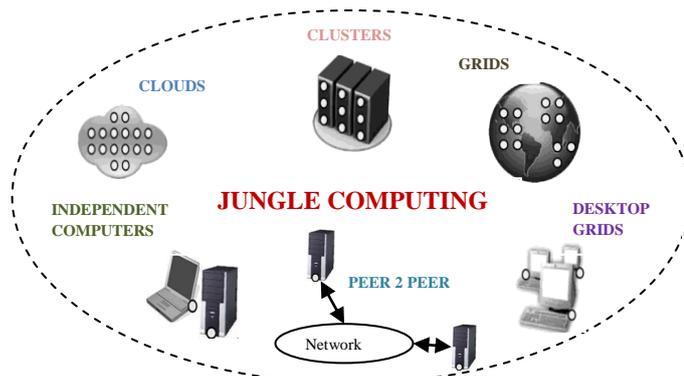

Figure 6.1: The jungle computing - a diverse collection of computing

The increasing complexity of the high performance computing environment has provided a bewildering range of choices beside traditional supercomputers and clusters. Scientists can now use grid and cloud infrastructures, in a variety of combinations along with traditional supercomputers - all connected via fast networks. And the emergence of many-core technologies such as GPUs, as well as supercomputers on chip within these environments has added to the complexity. Thus high performance computing can now use multiple diverse platforms and systems simultaneously, giving rise to the term "computing jungle". Ibis high-performance distributed programming system is an example of the jungle computing.

**Conclusion**

We have discussed the motivation for distributing computing. It will continue flourishing. There are so many topics which are going very hot in the research and development topics in both the academic and industry for many years to come. In above all the cloud computing is the recent topic which is under development by so many industrial giant like Google, EMC, Microsoft, Yahoo, Amazon, IBM, etc. This review paper enlightens the new paradigms of distributing computing. It will be beneficial for the students and the researchers.